\documentclass[twocolumn,showpacs,preprintnumbers,amsmath,amssymb]{revtex4}

\usepackage{graphicx}
\usepackage{epsfig}

\usepackage{longtable}
\usepackage{longtable}

\begin{document}

\title{8Dim calculations of the third barrier in $^{232}$Th and
        a conflict between theory and experiment on uranium nuclei.}

 \author{P.~Jachimowicz}
 \affiliation{Institute of Physics,
University of Zielona G\'{o}ra, Szafrana 4a, 65516 Zielona
G\'{o}ra, Poland}

\author{M.~Kowal} \email{m.kowal@fuw.edu.pl}
\affiliation{National Centre for Nuclear Research, Ho\.za 69,
PL-00-681 Warsaw, Poland}

\author{J.~Skalski}
\affiliation{National Centre for Nuclear Research, Ho\.za 69,
PL-00-681 Warsaw, Poland}

\date{\today}

\begin{abstract}
We find the height of the third fission barrier $B_{III}$ and energy of the
 third minimum $E_{III}$ in $^{232}$Th using the macroscopic - microscopic
 model, very well tested in this region of nuclei.
 For the first time it is done on an 8-dimensional deformation hypercube.
 The dipole distortion is included among the shape variables to assure that
 no important shapes are missed.
 The saddle point is found on a lattice containing more than 50 million
 points by the immersion water flow (IWF) method.
 The shallow third minimum, $B_{III}-E_{III}\approx 0.36$ MeV, agrees with
 experimetal data of Blons et al.
 This is in a sharp contrast with the status of the IIIrd minima in
 $^{232-236}$U: their experimental depth of $\geq3$ MeV contradicts
  all realistic theoretical predictions. We emphasize the importance
 of repeating the experiment on $^{232}$Th, by a technique similar to
 that used in the uranium nuclei, for settling the puzzle of the third minima
 in actinides.

\end{abstract}


\maketitle



 Search for a hyperdeformation (HD), i.e., extremely elongated
 shapes of atomic nuclei (spheroid with a major to minor axis
ratio of 3:1), is one of the most difficult challenges of modern
nuclear structure studies.
 There is still no convincing evidence for discrete gamma transitions of HD
 rotational bands.
 The only experimental evidence for hyperdeformation, except in light nuclei,
 has been reported in the region of light actinides, in particular
 in uranium isotopes $^{232}$U, $^{234}$ U and $^{236}$U, see
 \cite{Kra} and references therein.
 In these experiments, the fission
probability was studied as a function of the excitation energy via
different reactions. Strong resonances were observed and
 a fine structure of some interpreted as a signature of multiple
  hyperdeformed bands.
Measurements of the angular distribution of the fragments arising
from the induced fission support this interpretation. Moreover, by
measuring the angular distribution of the fission fragments one
can obtain some information about the spin and $K$ - quantum
number. Due to the predicted static mass-asymmetry of the third minima, their
 intrinsic parity would be broken, so
 the low-energy excitations should show a pattern of alternating parity
 (even nuclei) or parity doublet (odd nuclei) bands, different from that in
 the second minima. Since the third minima are predicted
 axially symmetric, the excitations would have an approximate (up to a
 Coriolis mixing, much weaker than in the first or second minimum)
 $K$ quantum number.

 The current experimental status of the third barrier in $^{232}$Th
 is different from that in uranium isotopes.
 No recent data exist, the only available are those from the prior 30 years
\cite{Blons1,Blons2,Blons3,Blons4}. The experiment conducted by
Blons et al pointed to a shallow third minimum (no
deeper than 0.5 MeV).
 Only a little more than 1 MeV deep third minimum
 in a neighboring nucleus $^{232}$Pa was recently reported by the
 Munich-Debrecen group \cite{Csige2012}.
 The technique used to resolve observed resonances was the same as
 in the study of uranium nuclei by the same group.
 Deep third minima in uranium ( $3 \div 4$ MeV) disagree with all calculations
 \cite{Mcdonell1,Mcdonell2,Bender,Gogny,HFBCS,Delar,MollerPRC2012}
 except those of S.~\'Cwiok et al, within the Woods-Saxon
 model \cite{Stefan,Stefan1,Stefan2}. Self-consitent calculations
based on the Skyrme SkM* interaction do not give a hyperdeformed
minimum in $^{232}$Th \cite{Mcdonell1} at all. With the help of a
temperature effect, a slight minimum appears, but does not exceed
500 keV. Moreover, with an increasing excitation, the depth of
this minimum decreases again. Situation is very similar in
$^{232}$U and $^{234}$U nuclei \cite{Mcdonell2}. Selfconsitent
calculations with the Gogny D1S interaction do not give third
minima in those nuclei \cite{Gogny,HFBCS,Delar} as well. Moreover,
for
 $^{236}$U, a recent result of M\"{o}ller et al
\cite{MollerPRC2012} within the macro-micro approach also does not
support a deep minimum reported in \cite{HyperU}: one can
read from the map (Fig. 8. in ref. \cite{MollerPRC2012}) that
 the minimum is not deeper than 300 keV,
  ten times smaller than the experimental value.

In fact, hyperdeformed minima predicted in \cite{Stefan,Stefan1,Stefan2}
 were quite abundant: they not only appeared in nuclei where nobody saw
them in experiment, but were double, with different octupole
 deformations $\beta_3\approx 0.3$ and 0.6. Using the same model,
 we have just showed in \cite{kowskalIIImin} that both types of the IIIrd minima
 in uranium nuclei disappear completely after proper nuclear shapes are considered.
 The more mass-asymmetric minima were unphysical from the beginning.
 Their quadrupole moment is large, $Q_{20} \simeq 170$ b, which situates
 them behind the barrier, closer to the scission point.
 The existence of minima with a smaller mass asymmetry (and quadrupole moments)
requires a more detailed study. Finding the barrier in
many-dimensional space requires hypercube calculations. This is the
case of $^{232}$Th, where the less mass-asymmetric minimum is the
deeper one.

  Previously, we found in \cite{kowskalIIImin} a tentative upper limit of
  330 keV for the IIIrd barrier in $^{232}$Th and 0 in even uranium isotopes
  $^{232-236}$U by 6Dim hypercube calculations and by probing various trial
  8Dim fission paths.
 The main aim of the present study is to ascertain the prediction
 for $^{232}$Th, showing that a third minimum is there.
 To accomplish this, an 8-dimensional grid calculation has
 been done.
 All deformations were treated on the same footing,
  without introducing any subdivision into relevant and irrelevant
  subspaces.
 Let us emphasize that till now only 5-dimensional macroscopic-microscopic
calculations of fission saddles were available in the literature
\cite{MollerNATURE2001,MollerPRL2004,MollerPRC2012}.

 There is one more important motivation for such studies. To
distinguish between resonances in the second and third minimum,
 the excitation energy should be higher than the first
barrier but simultaneously lower than the second one. We show that
  our multidimensional calculations predict such a possibility in $^{232}$Th,
   which is a natural candidate for the future experimental study.


 The energy is calculated within
 the microscopic-macroscopic method with a deformed Woods-Saxon potential.
 Nuclear shapes are defined in terms of the nuclear surface
 \cite{WS}. Since we found that nonaxiality in the region of
  the third barrier is less important than a proper treatment
  of the axially symmetric shapes, we parameterize the surface as
 \vspace{0.0cm}
  \begin{equation}
   \label{shape1}
    R(\theta)= c(\{\beta\})R_0 (1+\sum_{\lambda=1}\beta_{\lambda 0}
   Y_{\lambda 0}(\theta) )  ,
  \end{equation}
  where $c(\{\beta\})$ is the volume fixing factor.
The macroscopic energy is calculated using the Yukawa plus exponential
model \cite{KN} with parameters specified in \cite{WSpar}.
 All other parameters of the model are the
same as in a number of our previous studies, e.g. in
\cite{KOW10,kowskal,IIbarriers}.
 A hyperdeformed minimum was searched by exploring energy
 surfaces in a region of deformations beyond the second minimum.
 We generate the grid in axially symmetric deformations:
  \vspace{0.0cm}
   \begin{center}
\begin{eqnarray}
\beta_{10} & = &  \ \  -0.35 \ (0.05)\ 0.00  ; \,\,\,  \beta_{20}  =  0.55 \ (0.05) \ 1.50 ,    \nonumber\\
\beta_{30} & = &  \ \  0.00  \ (0.05)\ 0.35  ; \,\,\,   \beta_{40} =  -0.10 \ (0.05) \ 0.35 ,   \nonumber \\
\beta_{50} & = &  \ \  -0.20 \ (0.05) \ 0.20 ; \,\,\,  \beta_{60}  =  -0.15 \ (0.05) \ 0.15 ,   \nonumber \\
\beta_{70} & = &  \ \  -0.20 \ (0.05) \ 0.20 ; \,\,\,  \beta_{80}  =  -0.15 \ (0.05) \ 0.15.    \nonumber\\
\end{eqnarray}
\end{center}
Numbers in the parentheses specify the step with which the
calculation is done for a given variable. Thus, we finally have
the values of energy at a total of $50803200$ grid
points. On a such giant 8-dimensional grid the IWF
saddle point searching method is used.

\begin{figure}[hr]
\hspace{-2cm}
\begin{minipage}[hr]{52mm}
\centerline{\includegraphics[scale=0.33,angle=0.0]{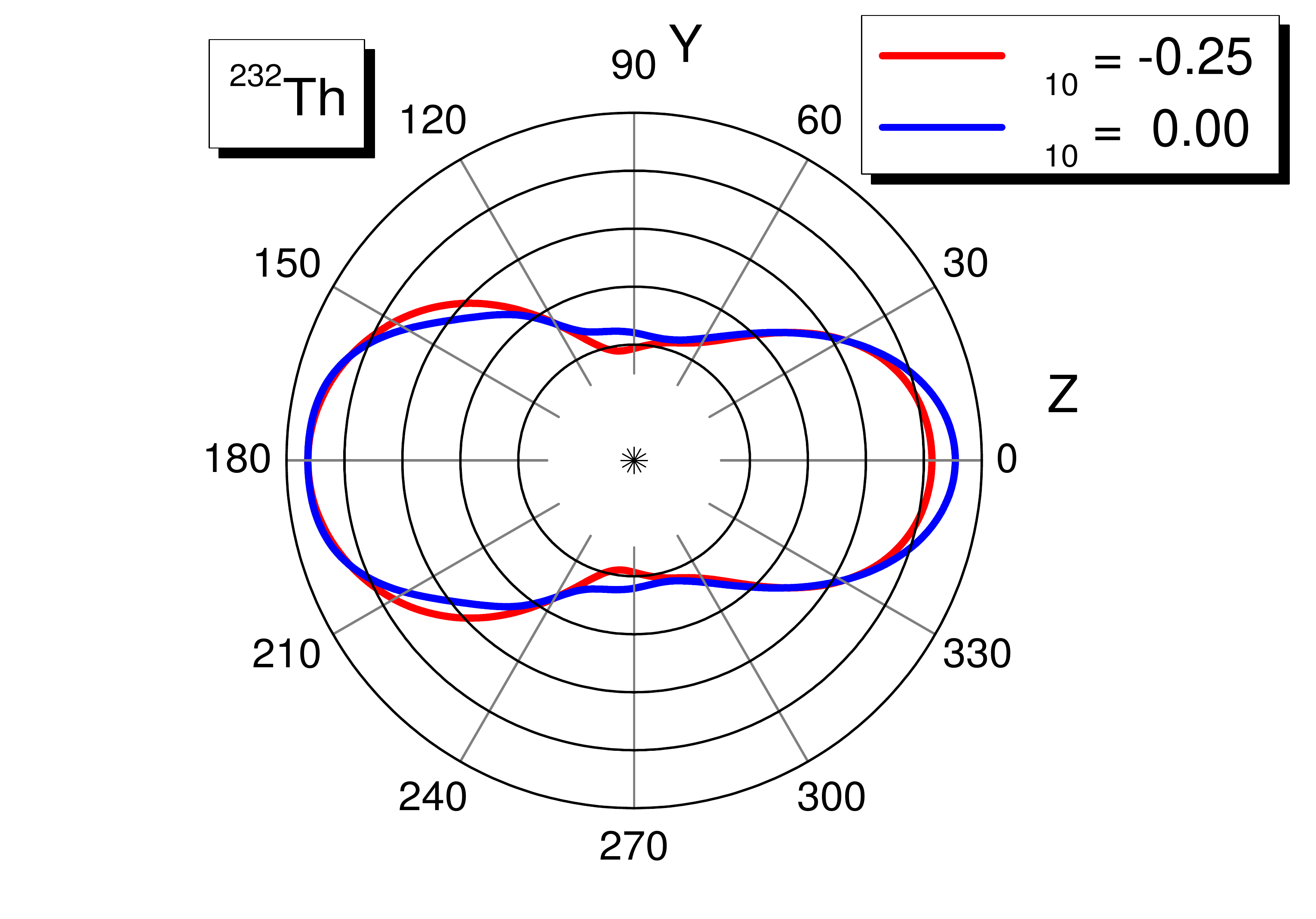}}
\end{minipage}
\caption{{\protect\small Shapes $R(\theta)/R_0)$ near to the IIIrd saddle:
 red line - 8D calculation ($\beta_{10}-\beta_{80}$), blue line
 - 7D calculation (without $\beta_{10}$).  }}
   \label{fig3}
\end{figure}

 Let us emphasize that our model, {\it without} the $\lambda=1$ term in
 (\ref{shape1}), very well reproduces first \cite{KOW10} and second barriers
 \cite{IIbarriers} and second minima \cite{kowskal} in actinide nuclei.
 One could hope that an extrapolation to more elongated shapes would
 be still valid within such a parametrization. What then causes troubles
 beyond the second peak?
 They result from the truncation of the expansion (\ref{shape1})
 that restricts possible shapes for large deformations.
 Two shapes close to the IIIrd saddle are shown in Fig. \ref{fig3}: one
 from the 8D calculation (nonzero $\beta_{10}$-$\beta_{80}$, in red)
 and the other, from the 7D calculation $\beta_{20}-\beta_{80}$ (in blue).
 Even if they seem not much different, the difference in barrier is 4 MeV.
 It appears that the dipole deformation $\beta_{10}$, usually associated only
 with a shift of the center of mass, effectively makes up for
 truncated multipoles in very deformed, mass-asymmetric configurations.
 We emphasize that our program keeps automatically the center of mass at
 the coordinate origin, thus $\beta_{10}$ serves only probing somewhat
 different shapes.
\begin{figure}[hr]
\hspace{-1cm}
\includegraphics[scale=0.43]{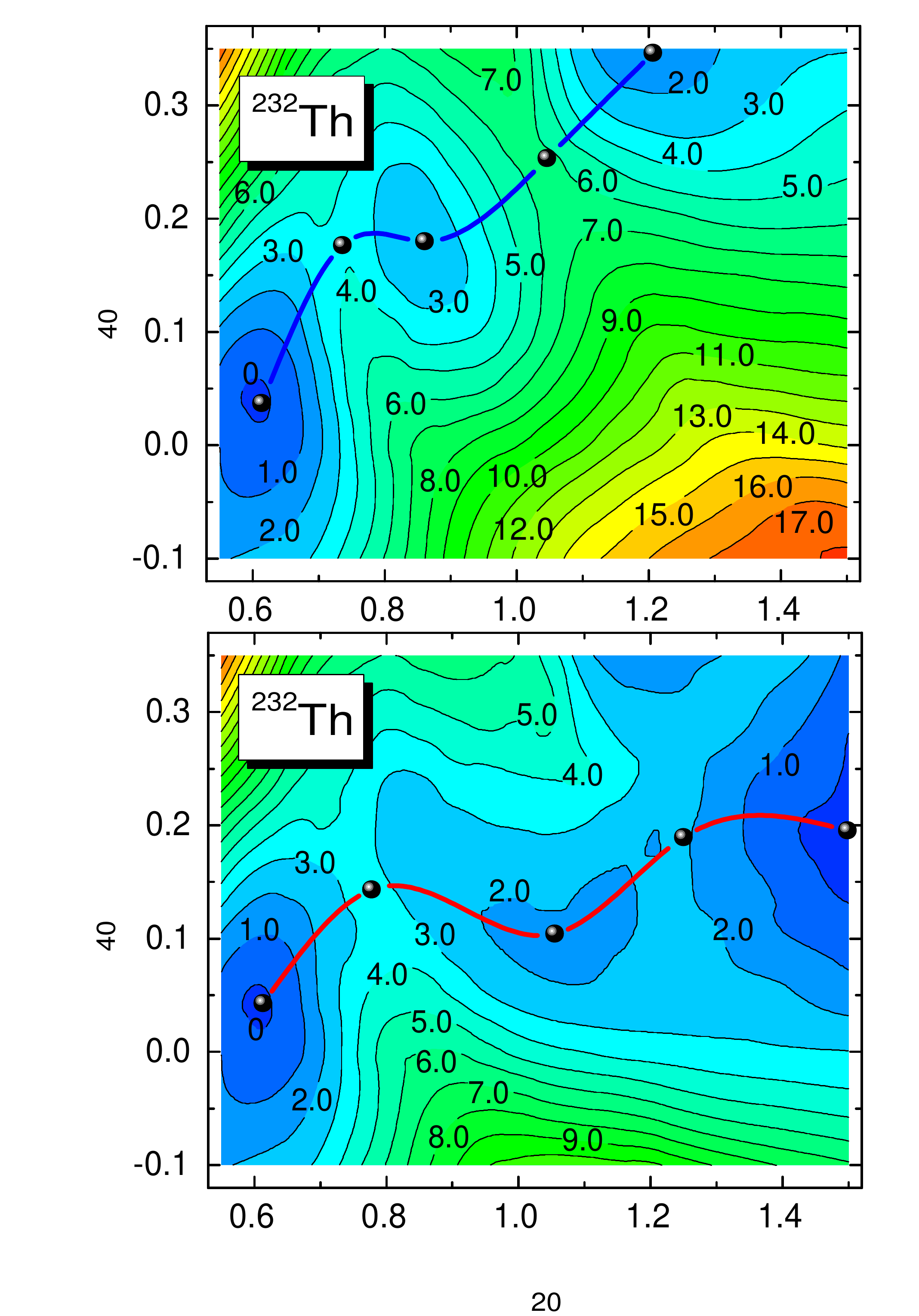}
\caption{{\protect\small
   Potential energy surfaces $E({\beta_{20},\beta_{40}})$ for $^{232}$Th from the 8D
  $\beta_{10}-\beta_{80}$ calculation, minimized over remaining degrees of freedom;
  upper panel - $\beta_{10}=0$, bottom panel - $\beta_1=min$.}}
\label{fig1}
\end{figure}

\begin{figure}[hr]
\begin{minipage}[hr]{52mm}
\centerline{\includegraphics[scale=0.23, angle=0.0]{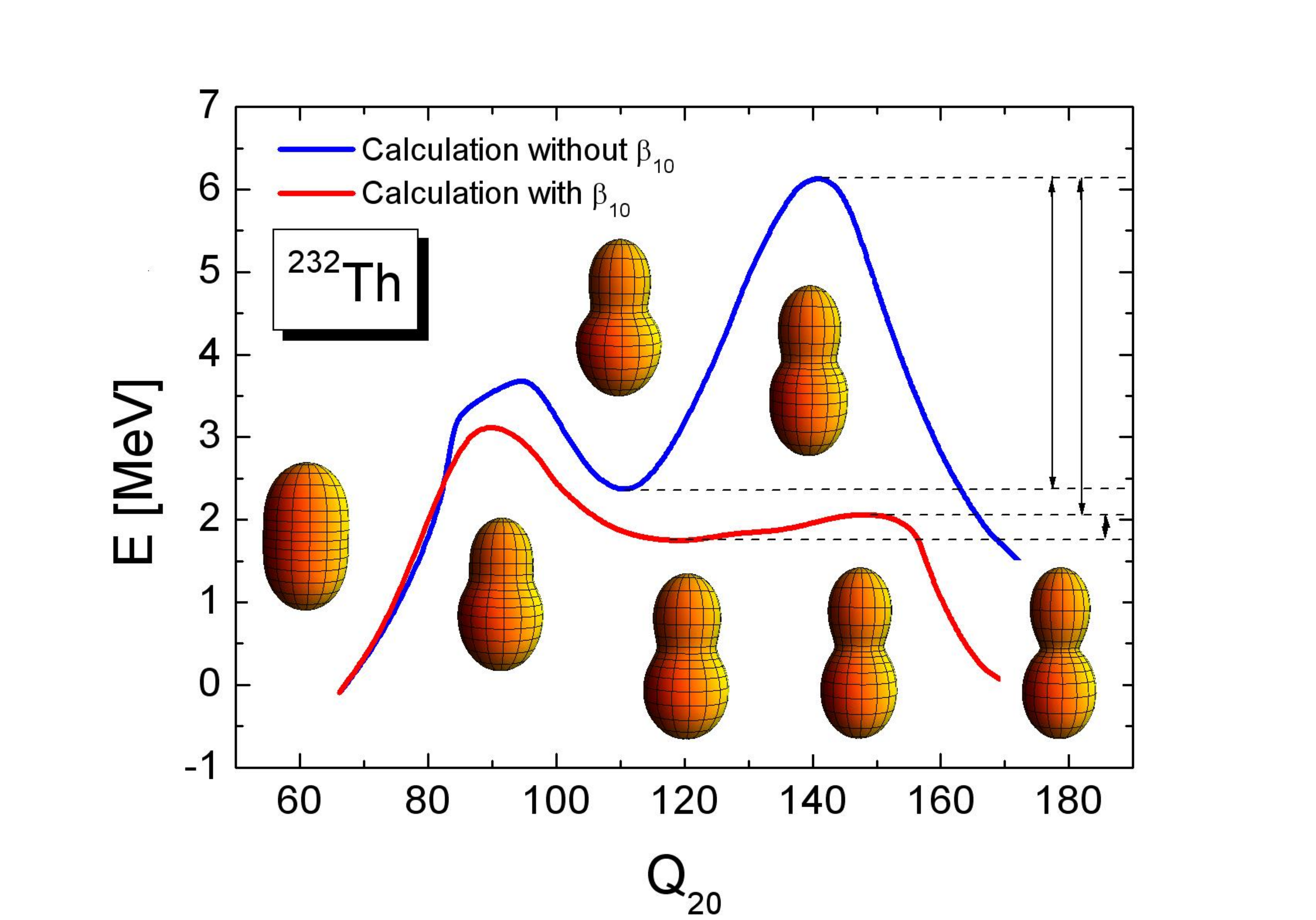}}
\end{minipage}
\caption{{\protect\small Energy along a sequence
  of smoothly elongating shapes, beginning at
   the superdeformed minimum. Appropriate shapes along trajectories are shown.
   }}
   \label{fig2}
\end{figure}

A modification of the energy landscape by including $\beta_{10}$ is
shown in Fig. \ref{fig1}. Since dipole is the first
 spherical harmonic (Eq.\ref{shape1}), one can
suspect its pronounced effect. Indeed, that effect is significant.
 Not only the height of the third saddle is clearly reduced, but the whole
 landscape changed. The result indicates a large effect of new shapes,
 unattainable previously. Figures show also examples of two fission paths.
In the version of calculation without the dipole (blue
trajectory), after passing through the second saddle, the nucleus
falls into a deep third minimum. To be split, it needs to tunnel
 through a more than 4 MeV barrier. In the case with included
$\beta_{10}$ (red trajectory), after passing the second barrier
nucleus can easily split. One can notice that $\beta_{10}$
deformation has almost no effect on the position and height of the
second saddle. Precise lowering of the third barrier by
$\beta_{10}$ may be read from Fig. \ref{fig2} as 4.1 MeV. Beside
the significant reduction, one can also see a change in the barrier shape.
 This rather dramatic change of the fission path in the
multidimensional deformation space has a direct effect
 on fission half-life.

\begin{table}
\caption{ Calculated and experimental features of the fission
barrier in $^{232}$Th and $^{232}$U. $M$ - ground state mass
excess, $B_{I}$ - first barrier, $E_{II}$- second minimum $B_{II}$
- second barrier $E_{III}$- third minimum $B_{III}$ - third
barrier.
All quantities (in MeV) are shown relative to the ground state.}
\label{1}
\begin{ruledtabular}
\begin{tabular}{ccc|cc|cccc}

 \multicolumn{1}{l}{Theory:} &
      \multicolumn{1}{c} {$M^{th}$}& {$B^{th}_{I}$} & {$E^{th}_{II}$}& {$B^{th}_{II}$} &{$E^{th}_{III}$} & {$B^{th}_{III}$} \\
      \multicolumn{1}{l}{Experiment:} & {$M^{exp}$}& {$B^{exp}_{I}$} & {$E^{exp}_{II}$}& {$B^{exp}_{II}$} &{$E^{exp}_{III}$} & {$B^{exp}_{III}$}\\ \hline

  $^{232}$Th                                  & 35.33   & 4.4          &    2.2   & 6.1         &     4.1   & 4.4      \\
  \cite{RIPL3}                                & 35.45   & 5.8          &    -     & 6.7 (6.2)   &      -    &  -      \\
  \cite{Blons2}                               & -       & 4.6          &    -     & 5.7         &     4.0   & 4.4      \\
 \hline
  $^{232}$U                                   & 34.34   & 4.5          &    3.2  & 5.7          &     0.0    & 0.0    \\
  \cite{RIPL3}                                & 34.61   & 5.4          &    -    & 5.4 (5.3)    &     -      & -      \\
  \cite{Hyper232U}                            & -       & 4.0          &    3.1  & 4.9          &     3.2    & 6.0    \\
\end{tabular}
\end{ruledtabular}
\end{table}

  A detailed information about the structure of the fission barrier in
  $^{232}$Th is collected in Table \ref {1}.
  The recent theoretical results and experimental data for
  $^{232}$U \cite{Hyper232U} are also given. It is seen that the model
  calculation well reproduces essential features  of the fission
  barrier up to superdeformed shapes in both nuclei.
  The inner fission barrier $B^{th}_{I}$ has been found according
  to the prescription presented in \cite{KOW10}.

   Here, a comment is needed. Usually the calculated first saddles
  are substantially higher than experimental values but the opposite
   situation takes place in the light Thorium isotopes
 what is called in the literature the ''Thorium anomaly''.
 In our case the nonaxiality reduces the first saddle for light Thorium
  only non-significantly, to a nearly proper height
 (see column 3 of Table \ref{1} column 3). Thus, the mentioned anomaly does
 not occur in our case (we use the modern data of \cite{Hyper232U}).

  In order to determine the second peak $B^{th}_{II}$, a mass asymmetry is
  included, details can be found in \cite{IIbarriers}. One can see
  that our calculated second minimum and second barrier reasonably agree with
 experiment. As already mentioned, the experimental situation
   (when it comes to the depth of hyperdeformed minima)
   in two isobars: $^{232}$Th and $^{232}$U is completely
  different. While in $^{232}$U, the IIIrd minimum is quite deep, in $^{232}$Th
  it is very shallow. Our results support the existence
 of a shallow third minimum in $^{232}$Th as in old experiments of Blons and
 thus agree with the majority of modern theoretical models.

 A designed experiment for $^{232}$Th, by a technique described in ref
\cite{Kra,Thirolf}, would be crucial for solving
 the mystery of the third minima in actinides.
 Particularly promising are experiments employing
highly monochromatic $\gamma$-ray beams for photofission studies.
With a high quality of photon and spectral intensity, exceeding the
performance of existing facilities by several orders of magnitude
seems to be possible in the near future \cite{Thirolfexp}.
 If the result of Blons et al is confirmed, we will have to understand
\emph{why between $^{232}$Th and $^{232}$U, two beta decays away,
  the energy landscape changes so dramatically}. On the other hand, if in the
future experiment, a depth of the IIIrd minimum
 is obtained similar as that in $^{232}$U, we will have a total contradiction
 between theory and experiment. \emph{Assuming that the existing
resonances cannot be interpreted otherwise, all current
meaningful theoretical models would have to be reconstructed anew to give
 deep hyperdeformed minima}.

 One can expect
that in hyperdeformed nuclei, some high particle states at normal deformation
 become occupied with increasing deformation. Since these are
 the orbits that are occupied in superheavy nuclei at moderate deformations,
 the whole question may have some impact on the understanding
 of superheavy nuclei.

In summary:

 (i) We have presented for the first time an 8D
hypercube calculation for $^{232}$Th. To find the third barrier on
a giant grid, the IWF method has been applied. After a proper
inclusion of the dipole deformation we found the depth of the
third minimum of about 0.36 MeV. This would rather exclude
spectroscopic studies in the IIIrd well.

(ii) Including a dipole distortion lowers the third saddle
 by more than 4 MeV. It seems likely that, with the shape parametrization
 (\ref{shape1}), the dipole deformation is important
 everywhere, where large elongation and necking is combined with a sizable mass
 asymmetry. For example, it may be the case of
 the Jacobi shape transition at high spins in medium-heavy nuclei.

(iii) New experimental study dedicated to hyperdeformation in
$^{232}$Th seems essential for the understanding of the third minima in
 actinide nuclei.

 {\bf Acknowledgments} This work was partially supported by Narodowe Centrum Nauki Grant
No. 2011/01/B/ST2/05131. The support of the LEA COPIGAL fund is
gratefully acknowledged.

\end{document}